\documentclass[aps,prd,twocolumn,showpacs]{revtex4}
\usepackage{graphicx}
\usepackage{dcolumn}
\usepackage{bm}
\begin{document}
\title{False Vacuum in the Supersymmetric Mass Varying Neutrinos Model}

\author{Ryo Takahashi} 
\email{takahasi@muse.sc.niigata-u.ac.jp}
\affiliation{Graduate School of  Science and Technology,
 Niigata University,  950-2181 Niigata, Japan}
\author{Morimitsu Tanimoto}
\email{tanimoto@muse.sc.niigata-u.ac.jp}
\affiliation{Department of Physics,
 Niigata University,  950-2181 Niigata, Japan}

\date{\today}

\begin{abstract}
We present detailed analyses of the vacuum structure of the scalar potential in a supersymmetric Mass Varying 
Neutrinos model. The observed dark energy density is identified with false vacuum energy and the dark energy 
scale of order $(10^{-3}\mbox{eV})^4$ is understood by gravitationally suppressed supersymmetry breaking scale, 
$F(\mbox{TeV})^2/M_{\mbox{{\scriptsize Pl}}}$, in the model. The vacuum expectation values of sneutrinos should 
be tiny in order that the model works. Some decay processes of superparticles into acceleron and sterile neutrino
 are also discussed in the model.
\end{abstract}


\maketitle

\section{Introduction}
In recent years, many cosmological observations suggest that the universe is flat and its expansion is 
accelerating\cite{Riess}. The most common explanation for the question of the origin of such an acceleration is 
to assume the existence of unknown dark energy component. 

A dynamical model, which relates a dark energy scalar with neutrinos, so called Mass Varying Neutrinos (MaVaNs) 
scenario, was proposed as a candidate for the dark energy\cite{Hung,Weiner}. In the context of this scenario 
there are a lot of works\cite{Kaplan,smavans2,smavans1}. Among these works, we consider a MaVaNs model in the 
supersymmetric theory.  
 
The MaVaNs scenario assumes an unknown scalar field so called ``acceleron'' and a sterile neutrino. They are 
gauge singlets under the gauge group of the standard model. In the extension of the scenario to the 
supersymmetry, these fields are embedded into chiral multiplets\cite{smavans2,smavans1}. Typical time evolutions 
of the neutrino mass and the equation of state parameter of the dark energy are presented in Ref.\cite{smavans1},
 where the acceleron and a sterile neutrino are embedded into the single chiral multiplet. In this model, the 
observed dark energy density is identified with the false vacuum energy and the dark energy scale of order 
$(10^{-3}\mbox{eV})^4$ is understood by gravitationally suppressed supersymmetry breaking scale, 
$F(\mbox{TeV})^2/M_{\mbox{{\scriptsize Pl}}}$. These works are based on assumptions that the imaginary part of 
the acceleron and vacuum expectation values of sneutrinos vanish.

In this paper, we present a modified supersymmetric MaVaNs model of Ref.\cite{smavans1} and detailed analyses of 
the vacuum structure of the scalar potential taking account of the finite imaginary part of the acceleron. Since 
the vanishing imaginary part of the acceleron should be dynamically realized, discussions based on the assumption
 in Ref.\cite{smavans1} are inadequate to analyse the vacuum structure of the potential. Actually, we find the 
global minimum is different from one in the case of the vanishing imaginary part of the acceleron. The dark 
energy is identified with the false vacuum energy and the dark energy scale is given by 
$\rho_{\mbox{{\scriptsize DE}}}^{1/4}\sim F(\mbox{TeV})^2/M_{\mbox{{\scriptsize Pl}}}$ in Ref.\cite{smavans1}. We
 show that the identification and the observed dark energy scale are also realized in the modified model where 
the acceleron is appropriately stabilized. Furthermore, we present cosmological discussions for the magnitude of 
the Yukawa coupling among the acceleron and sterile neutrinos which is assumed to be of order one in 
Ref.\cite{smavans1}. In this framework, the model with non-vanishing vacuum expectation values of sneutrinos are 
also discussed. We find that the magnitude of VEVs of sneutrinos are strictly constrained in the supersymmetric 
MaVaNs model. Then, two body decay processes of the superparticles are also studied.

In section 2, we present a supersymmetric MaVaNs model and detailed analyses of the vacuum structure of the 
scalar potential with finite imaginary part of the acceleron and vanishing VEVs of sneutrinos. In section 3, we 
discuss the case of non-vanishing VEVs of sneutrinos in the model and show constraints on the magnitude of VEVs. 
In section 4, two body decays of the superparticles into acceleron and sterile neutrino are investigated. 
The section 5 is devoted to the summary.

\section{Vacuum Structure of the Scalar Potential}
\subsection{Scalar potential for the acceleron}
Let us discuss vacuum structure in a supersymmetric mass varying neutrinos model. We consider the following 
superpotential $W$ in the MaVaNs model,
 \begin{eqnarray}
  W&=&m_DLA+M_DLR+\frac{\lambda_1}{6}A^3+\frac{\lambda_2}{2}AAR+\frac{M_A}{2}AA\nonumber\\
   & &+\frac{M_R}{2}RR,\label{1-1}
 \end{eqnarray}
where $L$ and $R$ are chiral superfields corresponding to the left-handed lepton doublet and the right-handed
 neutrino, respectively.\footnote{Other supersymmetric models have been proposed\cite{smavans2,smavans1}.} The 
$A$ is a gauge singlet chiral superfield.  The scalar and spinor components of $A$ are represented by $\phi$ and 
$\psi$, respectively. The scalar component corresponds to the acceleron which causes the present cosmic 
acceleration. The spinor component is a sterile neutrino. Coupling constants and mass parameters are represented 
by $\lambda_i(i=1,2)$ and $M_A$, $M_D$, $M_R$, $m_D$, respectively. The first and second terms of the right-hand 
side in Eq. (\ref{1-1}) are derived from Yukawa interactions such as $y_1HLA$ and $y_2HLR$ with 
$y_1\langle H\rangle=m_D$ and $y_2\langle H\rangle=M_D$, respectively, where $H$ is the usual Higgs doublet. In 
general, other interactions among gauge singlet fields such as $\lambda_1'R^3$ and $\lambda_2'AR^2$ are allowed 
but contributions from these interactions to the acceleron potential and neutrino masses are negligibly small. 
The terms appearing in Eq. (\ref{1-1}) are important to realize the MaVaNs scenario. In this model, there is no 
assignment to conserve the R-parity due to the emergence of trilinear coupling. In the limit of 
$y_1\rightarrow0$, a dark energy sector composed of $A$ is decoupled from the MSSM (visible) sector and then 
R-parity conservation is restored as in the MSSM. The LSP is absolutely stable in the case of conserved R-parity,
 however some decays of sparticles into the acceleron and a sterile neutrino are generally allowed in the model 
given by Eq. (\ref{1-1}). We will return to this point later.

Taking supersymmetry breaking effects into account, the scalar potential for the acceleron is given by
 \begin{eqnarray}
  V(\phi)&=&\frac{\lambda_1^2+\lambda_2^2}{4}|\phi|^4+(M_A^2+m_D^2-m^2)|\phi|^2\nonumber\\
         & &+\left(\frac{\lambda_1}{2}M_A|\phi|^2\phi
            -\frac{\kappa_1}{3}\phi^3+h.c.\right)\nonumber\\
         & &+\frac{\lambda_1}{2}m_D\tilde{\nu}_L\phi^{\dagger2}+m_DM_A\tilde{\nu}_L\phi^\dagger
            +m_DM_D\tilde{\nu}_R\phi^\dagger\nonumber\\
         & &+\frac{\lambda_2}{2}m_D\phi^\dagger\tilde{\nu}_L\tilde{\nu}_R^\dagger
            +\frac{\lambda_1\lambda_2}{4}|\phi|^2\phi\tilde{\nu}_R^\dagger\nonumber\\
         & &+\frac{\lambda_2}{2}M_A|\phi|^2\tilde{\nu}_R+\frac{\lambda_2}{2}M_D\tilde{\nu}_L\phi^{\dagger2}
            +\frac{\lambda_2}{2}M_R\tilde{\nu}_R\phi^{\dagger2}\nonumber\\
         & &-\kappa_2H\tilde{\nu}_L\phi
            -\kappa_3H\tilde{\nu}_R\phi+h.c.+V_0,
            \label{2}
 \end{eqnarray}
where $m$ and $\kappa_i(i=1\sim3)$ are supersymmetry breaking parameters and $V_0$ is a constant determined by 
imposing the condition that the cosmological constant is vanishing at the true minimum of the acceleron 
potential. 

In the Mass Varying Neutrinos scenario, the dark energy is supposed to be composed of the neutrinos and the 
scalar potential for the acceleron:
 \begin{equation}
  \rho _{\mbox{{\scriptsize DE}}}=\rho _\nu +V(\phi ),
 \end{equation}
where $\rho _{\mbox{{\scriptsize DE}}}$ and $\rho _\nu$ correspond to the energy densities of the dark energy and
 the neutrinos, respectively. For convenience later, we rewrite the scalar potential (\ref{2}) in terms of two 
real scalar fields $\phi_i$ ($i=1,2$) instead of $\phi$,
 \begin{eqnarray}
  V(\phi_i)
  &=&\frac{\lambda_1^2+\lambda_2^2}{4}(\phi_1^2+\phi_2^2)^2
     +\left(\frac{\lambda_1}{2}M_A-\frac{2}{3}\kappa_1\right)\phi_1^3\nonumber\\
  & &+\left(\frac{\lambda_1}{2}M_A+2\kappa_1\right)\phi_1\phi_2^2
     +\frac{m_\phi^2}{2}(\phi_1^2+\phi_2^2)\nonumber\\
  & &+(2M_A\tilde{\nu}_L+2M_D\tilde{\nu}_R+\lambda_2\tilde{\nu}_L\tilde{\nu}_R)m_D\phi_1\nonumber\\
  & &+\left(\frac{\lambda_1\lambda_2}{2}\phi_1+\lambda_2M_A\right)\tilde{\nu}_R(\phi_1^2+\phi_2^2)\nonumber\\
  & &+\{\tilde{\nu}_L(\lambda_1m_D+\lambda_2M_D)+\lambda_2M_R\tilde{\nu}_R\}(\phi_1^2-\phi_2^2)\nonumber\\
  & &-2(\kappa_2\tilde{\nu}_L+\kappa_3\tilde{\nu}_R)H\phi_1+V_0,\label{4}
 \end{eqnarray}
where we define as $\phi\equiv\phi_1+i\phi_2$ and $m_\phi^2\equiv2(M_A^2+m_D^2-m^2)$. 

First we discuss the vacuum structure of this scalar potential in the case of the vanishing vacuum expectation 
values of sneutrinos,
 \begin{eqnarray}
  V(\phi_i)
  &=&\frac{\lambda_1^2+\lambda_2^2}{4}(\phi_1^2+\phi_2^2)^2
     +\left(\frac{\lambda_1}{2}M_A-\frac{2}{3}\kappa_1\right)\phi_1^3\nonumber\\
  & &+\left(\frac{\lambda_1}{2}M_A+2\kappa_1\right)\phi_1\phi_2^2
     +\frac{m_\phi^2}{2}(\phi_1^2+\phi_2^2)+V_0.\nonumber\\
  & &\label{5}
 \end{eqnarray}
If $(\lambda_1M_A/2-2\kappa_1/3)<0$, the scalar potential has a local minimum at the origin in field space. 
Moreover, when parameters satisfy a relation,
 \begin{eqnarray}
  m_\phi<-\sqrt{\frac{2}{\lambda_1^2+\lambda_2^2}}\left(\frac{\lambda_1}{2}M_A-\frac{2}{3}\kappa_1\right),
 \end{eqnarray}
the potential has a second local minimum at $\phi_1>0$ and $\phi_2=0$, $V_2$, which is lower than the local 
minimum at the origin, $V_1$. Then, $V_2$ is given as follows:
 \begin{eqnarray}
  V_2\simeq-\frac{27}{4(\lambda_1^2+\lambda_2^2)^3}\left(\frac{\lambda_1}{2}M_A-\frac{2}{3}\kappa_1\right)^4.
 \end{eqnarray}
In a region of
  \begin{eqnarray}
  -\frac{(\lambda_1M_A+4\kappa_1)+M}{\lambda_1^2+\lambda_2^2}<\phi_1<
  -\frac{(\lambda_1M_A+4\kappa_1)-M}{\lambda_1^2+\lambda_2^2},
 \end{eqnarray}
$\phi_2$ has non-vanishing VEVs, 
 \begin{eqnarray}
  \langle\phi_2\rangle=\pm\sqrt{-\frac{(\lambda_1^2+\lambda_2^2)\phi_1^2+(\lambda_1M_A+4\kappa)\phi_1+m_\phi^2}
                               {\lambda_1^2+\lambda_2^2}},
 \end{eqnarray} 
where $M\equiv\sqrt{(\lambda_1M_A+4\kappa)^2-4(\lambda_1^2+\lambda_2^2)m_\phi^2}$. The potential energy of the 
local minimum in this region, $V_3$, is given by
 \begin{eqnarray}
  V_3\simeq-\frac{(\lambda_1M_A+4\kappa_1)^6}{3\cdot32^2\kappa_1^2(\lambda_1^2+\lambda_2^2)}.
 \end{eqnarray}
Since $V_3$ is smaller than $V_2$ in the case of $\lambda_1M_A\ll\kappa_1$, $V_3$ is the value of global minimum 
of this
 scalar potential. Therefore, we find
 \begin{eqnarray}
  V_0=|V_3|\simeq\frac{(\lambda_1M_A+4\kappa_1)^6}{3\cdot32^2\kappa_1^2(\lambda_1^2+\lambda_2^2)}.
 \end{eqnarray}
\begin{figure}[t]
\begin{center}
\includegraphics[width=0.9\linewidth]{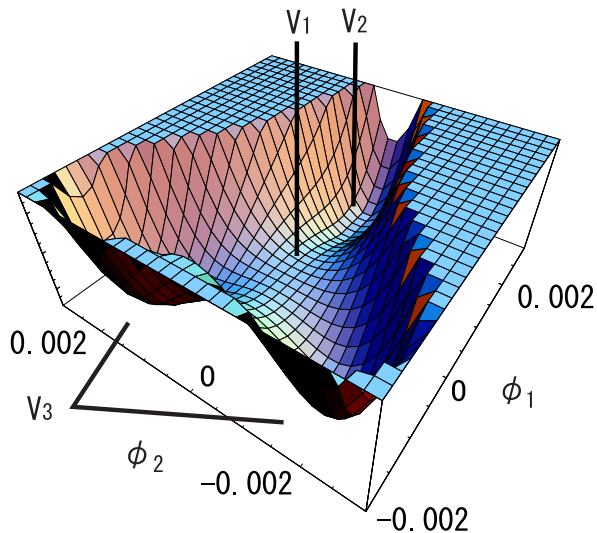}
\end{center}
\caption{The acceleron potential given in Eq. (\ref{5}). There are four local minima, the first one is the origin
 in field space, $V_1$, the second one is situated at $\phi_1>0$ and $\phi_2=0$, $V_2$, and the remaining two 
local minima are degenerated and global ones, $V_3$.}
\label{fig:pot}
\end{figure}
In summary, there are four local minima in the scalar potential (\ref{5}). They are shown in Fig.\ref{fig:pot}. 
The first one is the origin in field space and the second one is situated at $\phi_1>0$ and $\phi_2=0$. The 
remaining two local minima are degenerated and global ones of the scalar potential,
 \begin{eqnarray}
  V_3&\simeq& V(\phi_2=\pm\langle\phi_2\rangle_g),
 \end{eqnarray}
where
 \begin{eqnarray}
  \langle\phi_2\rangle_g&=&\sqrt{\frac{-(\lambda_1^2+\lambda_2^2)\phi_1^2-(\lambda_1M_A+4\kappa)\phi_1-m_\phi^2}
                               {\lambda_1^2+\lambda_2^2}}.\nonumber\\
  & &
 \end{eqnarray}
We find that a constant $V_0$ in Eq. (\ref{2}), which cancels the cosmological constant, is tuned such as 
$V_0=|V_3|$.

\subsection{Numerical analysis}
Now, let us consider a possible Mass Varying Neutrinos model with this scalar potential. Since the dark energy 
in the MaVaNs scenario is the sum of the scalar potential and the energy density of neutrinos, the stationary 
condition is given by
 \begin{eqnarray} 
  \frac{\partial\rho _{\mbox{{\scriptsize DE}}}}{\partial\phi_i}
  =\frac{\partial\rho_\nu}{\partial\phi_i}+\frac{\partial V(\phi_i)}{\partial\phi_i}=0.\label{11}
 \end{eqnarray}
We find the local minimum at the origin is slightly deviated due to the contribution of the neutrino energy 
density. And a numerical calculation shows that the model works near the origin in field space: 
$\langle\phi_1\rangle^0\simeq0$ eV and $\langle\phi_2\rangle^0=0$ eV, where superscript zero means the value at 
the present epoch. Since $\phi$ field is trapped in the local minimum near the origin, one observes the potential
 energy $|V_3|$ as the dark energy at the present epoch. The $\phi$ field trapped near the origin is plausible 
because it is likely that the universe starts at the origin due to the thermal effects. Typical values of 
parameters are given as 
 \begin{eqnarray}
  &&\kappa_1\simeq2.17\times10^{-3}\mbox{eV},\hspace{2mm}M_A\simeq1.003\times10^{-2}\mbox{eV},\nonumber\\ 
  &&|M_D^2/M_R|\simeq4.998\times10^{-2}\mbox{eV}, \hspace{2mm}m\simeq1.008\times10^{-2}\mbox{eV},\nonumber\\
  &&\label{14}
 \end{eqnarray}
by numerical calculation when we take
 \begin{eqnarray}
  &&m_{\nu_L}^0=5\times10^{-2}\mbox{eV},\hspace{5mm}m_{\psi}^0=10^{-2}\mbox{eV},\nonumber\\
  &&m_D=10^{-3}\mbox{eV},
    \hspace{5mm}m_\phi=10^{-4}\mbox{eV},\nonumber\\
  &&\lambda_1=10^{-9},\hspace{5mm}\lambda_2=1,\label{15-1}
 \end{eqnarray}
as inputs masses and couplings. Now, we assume that supersymmetry is broken at TeV scale in a hidden sector and 
its effects are transmitted to a dark energy sector, which includes a chiral superfield $A$, only through 
gravitational interaction. Then, the scale of soft terms of order 
(TeV)$^2/M_{\mbox{{\scriptsize pl}}}\sim\mathcal{O}(10^{-2}$-$10^{-3}\mbox{ eV})$ is expected. Soft parameters 
$\kappa_2$ and $\kappa_3$ are three-scalars couplings such as $\kappa_2\tilde{\nu}_LH\phi$ and 
$\kappa_3\tilde{\nu}_RH\phi$. If the dark energy sector is communicated with the visible sector, which includes 
the right-handed sneutrino, only through the gravitational interaction except for a peculiar Yukawa interaction 
of the MaVaNs scenario such as $y_1H\nu_L\psi$, $\kappa_{2,3}$ are roughly estimated of order 
$m_{\mbox{{\scriptsize soft}}}^2/M_{\mbox{{\scriptsize pl}}}$ by dimensional analysis, where 
$m_{\mbox{{\scriptsize soft}}}$ means the scale of soft terms in the visible sector. The magnitude of soft 
parameters in the visible sector depends on a mediation mechanism between the visible and a hidden sector. If we
 take $m_{\mbox{{\scriptsize soft}}}\sim\mathcal{O}(\mbox{TeV})$, the magnitude of $\kappa_{2,3}$ becomes 
$\mathcal{O}(10^{-3}\mbox{ eV})$. A larger scale of $\kappa_{2,3}$, which corresponds to
$m_{\mbox{{\scriptsize soft}}}$, prefers tiny or vanishing VEVs of sneutrinos in the MaVaNs model. We will return
 to this point later. 

The parameter $M_A$ affects on the neutrino masses. The effective neutrino mass matrix is given by 
 \begin{eqnarray}
  \mathcal{M}\simeq
   \left(
   \begin{array}{cc}
    c+\epsilon_1   & m_D+\epsilon_2 \\
    m_D+\epsilon_2 & M_A+\lambda_1\langle\phi_1\rangle+\epsilon_3
   \end{array}
  \right),
  \label{MM}
 \end{eqnarray}
in the basis of $(\nu_L,\psi)$, where $\nu_L$ and $\psi$ are the left-handed and a sterile neutrino, 
respectively. The $c$ in the $(1,1)$ element of this matrix corresponds to the usual term given by the seesaw 
mechanism in the absence of the acceleron, $c\equiv -M_D^2/M_R$, where 
$\lambda_1\langle\phi_1\rangle\ll M_D\ll M_R$ is assumed. The parameters $\epsilon_i(i=1\sim3)$ are functions of 
$\phi_1$, which are suppressed by large scales such as the right-handed neutrino mass scale $M_R$ or the higgsino
 mass $m_{\tilde{h}}$. If the VEV of $\phi_1$ and the mixing between the active and a sterile neutrino are small,
 the mass of a sterile neutrino is almost determined by $M_A$. We assume that the magnitude of a sterile neutrino
 mass is the same order as the left-handed neutrino one, $m_\psi^0=10^{-2}$ eV, and notice 
$m_\psi^0\simeq M_A\simeq 1.003\times10^{-2}$ eV as shown in Eqs. (\ref{14}) and (\ref{15-1}). 

The parameter $m_D$ determines the mixing between the active and sterile neutrinos. When we take as 
$m_D=10^{-3}$eV, the mixing angle, $\sin^2\theta\simeq4\times 10^{-4}$, is expected. The active and sterile 
neutrinos mix maximally in the case of $10^{-3}$eV $\ll m_D$, which is ruled out by the current astrophysical, 
cosmological and laboratory bounds.\footnote{Bounds on the mixing between active and sterile neutrinos are 
summarized in Ref. \cite{Smirnov}.} Since $m_D$ is defined as $y_1\langle H\rangle$ in our model, 
$m_D$ \raisebox{0.4ex}{$<$}\hspace{-0.75em}\raisebox{-.7ex}{$\sim$} $10^{-3}$eV and $\langle H\rangle\simeq10^2$ 
GeV require $y_1$ \raisebox{0.4ex}{$<$}\hspace{-0.75em}\raisebox{-.7ex}{$\sim$} $10^{-14}$. In general, MaVaNs 
models need such a tiny Yukawa coupling among the left-handed neutrino, the Higgs boson and the sterile neutrino.
 In the limit of $y_1\rightarrow0$, a dark energy sector is decoupled from the visible sector and the mixing 
between the left-handed and sterile neutrinos vanishes. Such a small Yukawa coupling may be explained in a brane
 model however the discussion is beyond the scope of this paper. We use the value $m_D=10^{-3}$eV, which 
corresponds to $y_1\simeq10^{-14}$, in numerical analyses.

In the MaVaNs scenario, the mass of the scalar field should be less than $\mathcal{O}(10^{-4}\mbox{eV})$ in order
 that the scalar field does not vary significantly on distance of inter-neutrino spacing, $n_\nu^{-1/3}$. 
Therefore, $m_\phi=10^{-4}$eV is taken in our calculation.

Next, we comment on couplings $\lambda_i$. The size of $\lambda_1$ is constrained by the successful Big Bang 
Nucleosynthesis (BBN) theory and supernova data. In the early universe, the acceleron and the sterile neutrino 
have to decouple in order that the scenario of BBN is not radically changed. If the acceleron production rates 
through acceleron-strahlung and pair annihilation of neutrinos are less than the expansion rate at the era of 
BBN, these particles are not in thermal equilibrium at the epoch\cite{Weiner}. These considerations give us
 \begin{eqnarray}
  &&\left(\frac{\partial m_\nu}{\partial\phi}\right)^2\leq1,\label{15}\\
  &&\left(\frac{\partial m_\nu}{\partial\phi}\right)^4<10^{-22},\hspace{1mm}
    \left(\frac{\partial^2m_\nu}{\partial\phi^2}\right)^2<10^{-34}\mbox{eV}^{-2}.\label{16-1}
 \end{eqnarray}
Eqs. (\ref{15}) and (\ref{16-1}) show constraints that $\phi$-production rates through $\phi$-strahlung and 
neutrino pair annihilation are smaller than the expansion rate at the BBN epoch, respectively.

Similar constraints come from consideration in supernovae. We require that $\phi$-emission does not cool a 
protoneutron star too quickly to change the observed neutrino spectra emitted in the first ten seconds. This 
requirement is realized in the following conditions
 \begin{eqnarray}
  &&\left(\frac{\partial m_\nu}{\partial\phi}\right)^2\leq10^{-12},\label{17}\\
  &&\left(\frac{\partial m_\nu}{\partial\phi}\right)^4<10^{-23},\hspace{1mm}
    \left(\frac{\partial^2m_\nu}{\partial\phi^2}\right)^2<10^{-37}\mbox{eV}^{-2}.\label{18}
 \end{eqnarray}
We find that constraints from supernova data, Eqs. (\ref{17}) and (\ref{18}), are more severe than the ones from 
the BBN scenario, Eqs. (\ref{15}) and (\ref{16-1}).

\begin{figure}[t]
\includegraphics[width=0.9\linewidth]{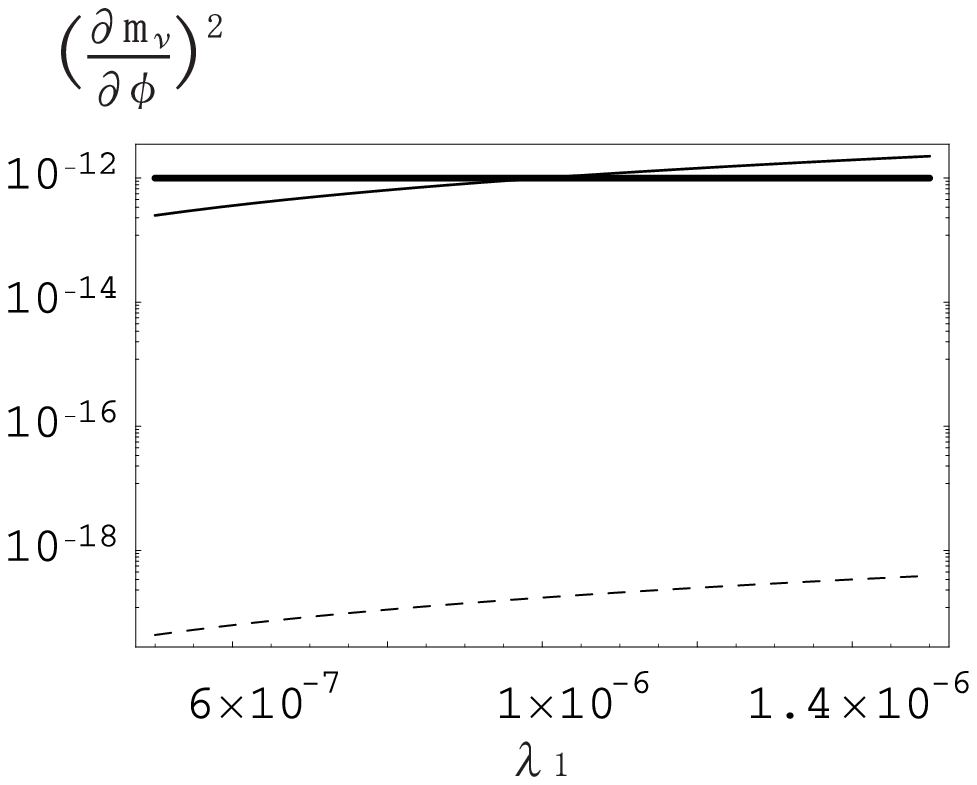}
\includegraphics[width=0.9\linewidth]{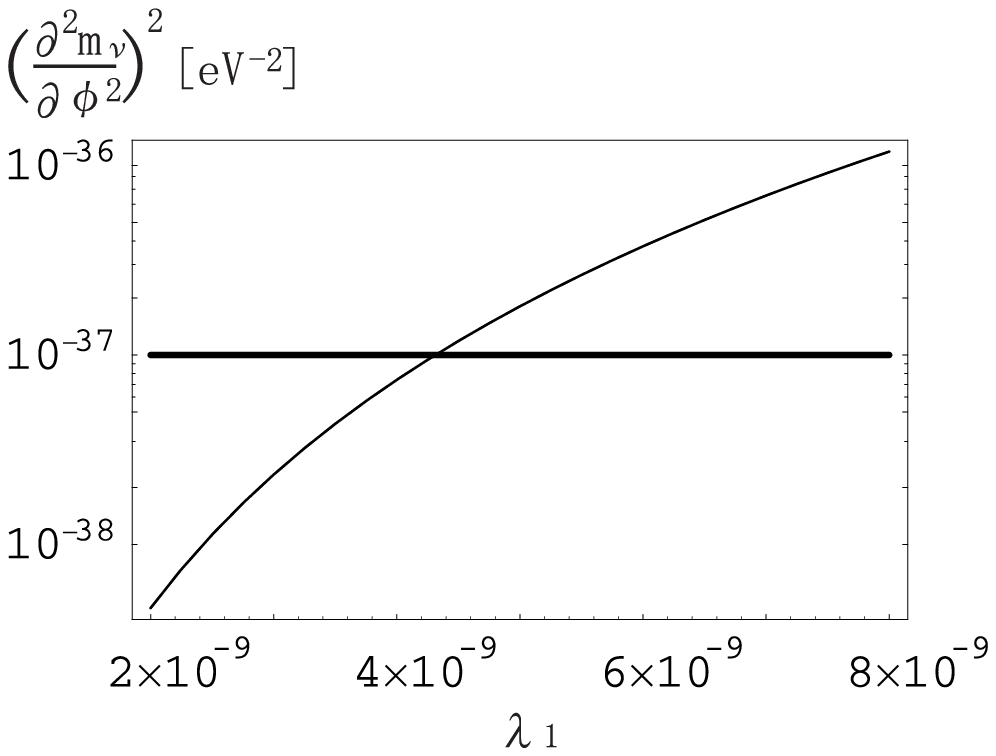}
\caption{The acceleron dependences of neutrino masses and the constraint on $\lambda_1$ from supernova data. The 
left figure shows the relation between $(\partial m_\nu/\partial\phi)^2$ and $\lambda_1$. The horizontal line 
shows the constraint of Eq (\ref{17}). The dashed and thin curves correspond to 
$(\partial m_{\nu_L}/\partial\phi_1)^2$ and $(\partial m_\psi/\partial\phi_1)^2$, respectively. The lower region 
than the horizontal line, $(\partial m_\nu/\partial\phi_1)^2<10^{-12}$, is allowed by the supernova data. In the 
right figure, the horizontal line shows the constraint of Eq. (\ref{18}). The thin curve corresponds to 
$(\partial^2m_{\nu_L}/\partial\phi_1^2)^2$, which is exactly equal to $(\partial^2m_\psi/\partial\phi_1^2)^2$. 
The lower region than the horizontal line, $(\partial^2m_\nu/\partial\phi^2)^2<10^{-37}$ eV$^{-2}$, is allowed by
 the supernova data. These figures mean that if $\lambda_1$ is smaller than about $4\times10^{-9}$, all 
constraints are satisfied.}
\label{fig:0}
\end{figure}

In our model, the acceleron dependences of masses of the left-handed and sterile neutrinos are given by 
diagonalizing the effective neutrino mass matrix (\ref{MM}),
 \begin{eqnarray}
  m_{\nu _L}(\phi_1)&=&\frac{c+M_A+\lambda_1\langle\phi_1\rangle}{2}\nonumber\\
                    & &+\frac{\sqrt{[c-(M_A+\lambda_1\langle\phi_1\rangle)]^2+4m_D^2}}{2},\\
  m_\psi(\phi_1)    &=&\frac{c+M_A+\lambda_1\langle\phi_1\rangle}{2}\nonumber\\
                    & &-\frac{\sqrt{[c-(M_A+\lambda_1\langle\phi_1\rangle)]^2+4m_D^2}}{2},
 \end{eqnarray}
where we ignore $\epsilon_i$. Fig.\ref{fig:0} shows that the acceleron dependences of neutrino masses, 
$(\partial m_\nu/\partial\phi_1)^2$ and $(\partial^2m_\nu/\partial\phi_1^2)^2$, and the constraint on $\lambda_1$
 from supernovae. Since the neutrino masses depend on only $\phi_1$, $(\partial/\partial\phi)$ is replaced with 
$(\partial/\partial\phi_1)$. The left and right figures in Fig.\ref{fig:0} corresponds to the constraint in Eq. 
(\ref{17}) and the second one in Eq. (\ref{18}), $(\partial^2m_\nu/\partial\phi^2)^2<10^{-37}$ eV$^{-2}$. We find
 that both constraints are satisfied if 
$\lambda_1$ \raisebox{0.4ex}{$<$}\hspace{-0.75em}\raisebox{-.7ex}{$\sim$} $4\times10^{-9}$. 

Next, we consider the coupling $\lambda_2$. In numerical calculations, we take the same number density of sterile
 neutrinos as the one of the left-handed neutrinos, $n_\nu\propto T^3$, where $T$ is the neutrino temperature. 
This means that sterile neutrinos should be thermalized in the early universe. The interaction rate of a sterile
 neutrino through the interaction $(\lambda_2/2)ARR$ is typically given by $\Gamma\sim\lambda_2^2T$. The 
thermalization of sterile neutrinos, $\Gamma>H(\sim T^2/M_{\mbox{{\scriptsize Pl}}})$, requires  
$\lambda_2^2M_{\mbox{{\scriptsize Pl}}}>T$. For instance, if we assume that sterile neutrinos thermalized after 
the inflation, $\lambda_2^2M_{\mbox{{\scriptsize Pl}}}>T_{RH}$ should be realized, where $T_{RH}$ is the 
reheating temperature. The thermalization of sterile neutrinos in this MaVaNs model with the coupling 
$\lambda_2$ of order one is easily realized by many inflation models.

\subsection{False vacuum}

The presented MaVaNs model works at the false vacuum. We estimate the decay probability of the false vacuum. The 
decay probability is given by $e^{-B}$, where $B$ is the action for the classical solution in Euclidean space. We
 use the result of \cite{duncan},
 \begin{eqnarray}
  B=2\pi^2\frac{(\Delta\phi)^4}{[(\Delta V_-)^{1/3}-(\Delta V_+)^{1/3}]^3},\label{23}
 \end{eqnarray}
where $\Delta\phi\equiv\phi_--\phi_+$, $\phi_+$ and $\phi_-$ are the vacuum expectation values at the false and 
the true vacuum, respectively, and
 \begin{eqnarray}
  \Delta V_\pm\equiv V_T-V(\phi_\pm),
 \end{eqnarray}
where $V_T$ is the potential energy of the top of the barrier between the false and the true vacuum. After 
calculating the dark energy density numerically in our model, we find
 \begin{eqnarray}
  &&\phi_+\simeq3.61\times10^{-6}\mbox{eV},\hspace{4mm}\phi_-\simeq2.83\times10^{-3}\mbox{eV},\nonumber\\
  &&\Delta V_+\simeq2.3\times10^{-20}\mbox{eV}^4,\hspace{4mm}\Delta V_-\simeq2.95\times10^{-11}\mbox{eV}^4,
    \nonumber\\
  &&
 \end{eqnarray}
and thus $B\simeq 42.9\gg 1$. Therefore, the decay probability is sufficiently small. This probability is of the 
decay from the false vacuum near the origin in the field space, $V_1$, into one of two degenerated true vacua, 
$V_{3}$. There is another possibility of the false vacuum decay, which is the one into another false vacuum, 
$V_2$. In this case, we get $B\simeq 41.1$, where we take $\phi_-\simeq2.80\times10^{-3}$ eV. In this numerical 
calculation, we take the square potential approximation in Ref. \cite{duncan}. In the limit of 
$\Delta V_+\rightarrow 0$, the result of this approximation Eq. (\ref{23}) describes the sheer drop which is the extreme case of ``tunnelling without barriers'' discussed in Ref. \cite{Lee}, 
 \begin{eqnarray}
  B=2\pi^2\frac{(\Delta\phi)^4}{\Delta V_-}.
 \end{eqnarray}
Since $\Delta V_+\ll\Delta V_-$ in our model, this approximation is appropriate rather than ``triangle'' 
\cite{duncan} or ``thin-wall'' approximation \cite{coleman}.

\section{Vacuum expectation values of sneutrinos}
In the previous section, we consider the vanishing vacuum expectation values of sneutrinos. Let us consider 
effects of non-vanishing VEVs of sneutrinos. VEVs of sneutrinos interacting with the acceleron are strictly 
constrained in the Mass Varying Neutrinos model. The most severe one comes from the stationary condition Eq. 
(\ref{11}). We rewrite the condition Eq. (\ref{11}) as
 \begin{eqnarray}
  \frac{\partial V}{\partial\phi_i}
  =-T^3\sum_{I=\nu_L,\psi}\frac{\partial m_I}{\partial\phi_i}\frac{\partial F(\xi_I)}{\partial\xi_I}\label{13},
 \end{eqnarray}
where $\rho_{\nu}=T^4\sum F(\xi_I)$, $\xi_I\equiv m_I/T$ and 
 \begin{eqnarray}
  F(\xi_I)\equiv\frac{1}{\pi^2}\int_0^\infty\frac{dyy^2\sqrt{y^2+\xi_I^2}}{e^y+1}.
 \end{eqnarray}
Since the right hand side in Eq. (\ref{13}) is approximately proportional to the number density of neutrinos and 
both active and sterile neutrino masses derived from the mass matrix Eq. (\ref{MM}) do not depend on $\phi_2$, 
the stationary condition at the present epoch is approximated by
 \begin{eqnarray}
  \left.\frac{\partial V}{\partial\phi_1}\right|_{\phi_1=\langle\phi_1\rangle^0}
  \simeq -n_\nu^0
         \sum_{I=\nu_L,\psi}\left.\frac{\partial m_I}{\partial\phi_1}\right|_{\phi_1=\langle\phi_1\rangle^0.}
  \label{16}
 \end{eqnarray}
A viable MaVaNs model should satisfy this constraint. Of course, the model with the vanishing VEVs of sneutrinos 
satisfies it. When we take $\lambda_1=10^{-9}$, we have
 \begin{eqnarray}
  &&\left(\frac{\partial m_{\nu_L}}{\partial\phi_1}\right)^2\sim\mathcal{O}(10^{-25}),\\
  &&\left(\frac{\partial m_\psi}{\partial\phi_1}\right)^2\sim\mathcal{O}(10^{-18}).
 \end{eqnarray}
Therefore, the magnitude of the right hand side in Eq. (\ref{16}) is $\mathcal{O}(10^{-22}\mbox{eV}^3)$, where we
 take $n_\nu^0\sim\mathcal{O}(10^{-13}\mbox{eV}^3)$. Since the term $m_\phi^2\phi_1$ is dominant in the left hand
 side of Eq. (\ref{16}), we have $\langle\phi_1\rangle^0\sim\mathcal{O}(10^{-14}\mbox{eV})$, where we take 
$m_\phi=10^{-4}$ eV.

When the vacuum expectation values of sneutrinos are non-vanishing, the following terms are strictly constrained 
unless miraculous cancellation among each terms is realized,
 \begin{eqnarray}
  \frac{\partial V(\phi_1)}{\partial\phi_1}
  &\supset&(2M_A\tilde{\nu}_L+2M_D\tilde{\nu}_R+\lambda_2\tilde{\nu}_L\tilde{\nu}_R)m_D\nonumber\\
  &       &+\lambda_1\lambda_2\tilde{\nu}_R\phi_1^2+2\lambda_2M_A\tilde{\nu}_R\phi_1\nonumber\\
  &       &+2\{\tilde{\nu}_L(\lambda_1m_D+\lambda_2M_D)+\lambda_2M_R\tilde{\nu}_R\}\phi_1\nonumber\\
  &       &-2(\kappa_2\tilde{\nu}_L+\kappa_3\tilde{\nu}_R)H.
 \end{eqnarray}
Especially, restrictions on the VEVs of sneutrinos are obtained from 
$\kappa_2H\tilde{\nu}_L$ and $\kappa_3H\tilde{\nu}_R$. In order that these terms do not spoil the model, 
$\langle\tilde{\nu}_{L,R}\rangle<10^{-39}$ eV must be realized when we take $\kappa_{2,3}=10^{-3}$eV and 
$\langle H\rangle=10^2$GeV. The larger values of $\kappa_{2,3}$ gives the smaller VEVs of sneutrinos. The 
vanishing or tiny VEVs of sneutrinos are favored in this supersymmetric Mass Varying Neutrinos model.

\section{Sparticle decay}
The Mass Varying Neutrinos scenario assumes some light particles such as the acceleron and the sterile neutrino. 
In the supersymmetric realization as we discussed, these particles are embedded into single chiral multiplet. 
Since Eq. (\ref{1-1}) leads to R-parity violation in order that a given scalar potential and the neutrinos mass 
are consistent with the current cosmological observations, some heavy sparticle such as the sneutrino and the 
higgsino decay into light particles due to such R-parity violating interactions.

\begin{figure}[t]
\includegraphics[width=0.9\linewidth]{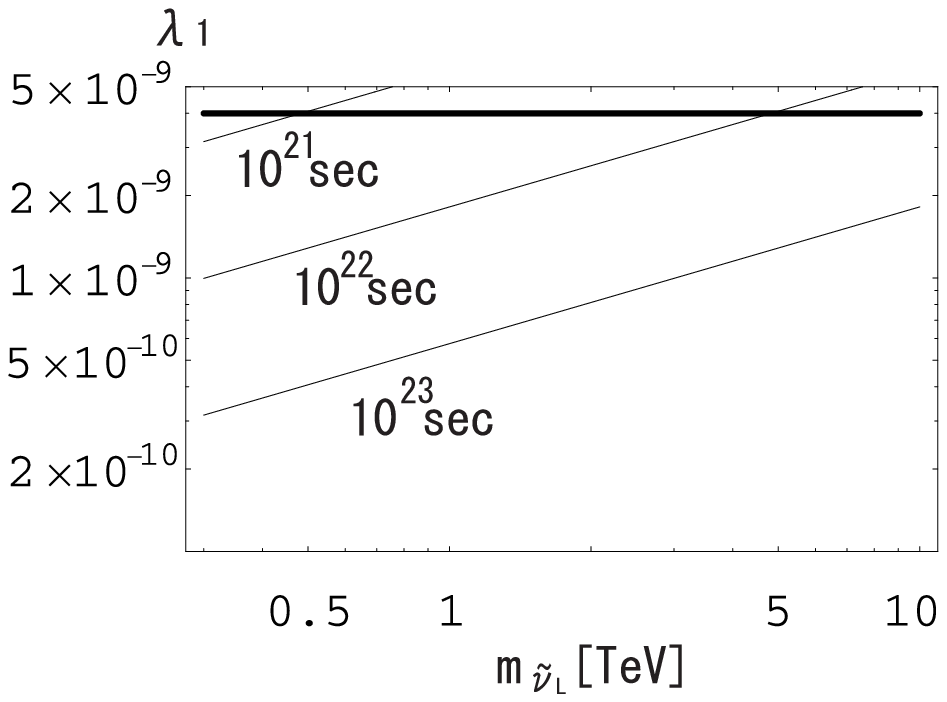}
\includegraphics[width=0.9\linewidth]{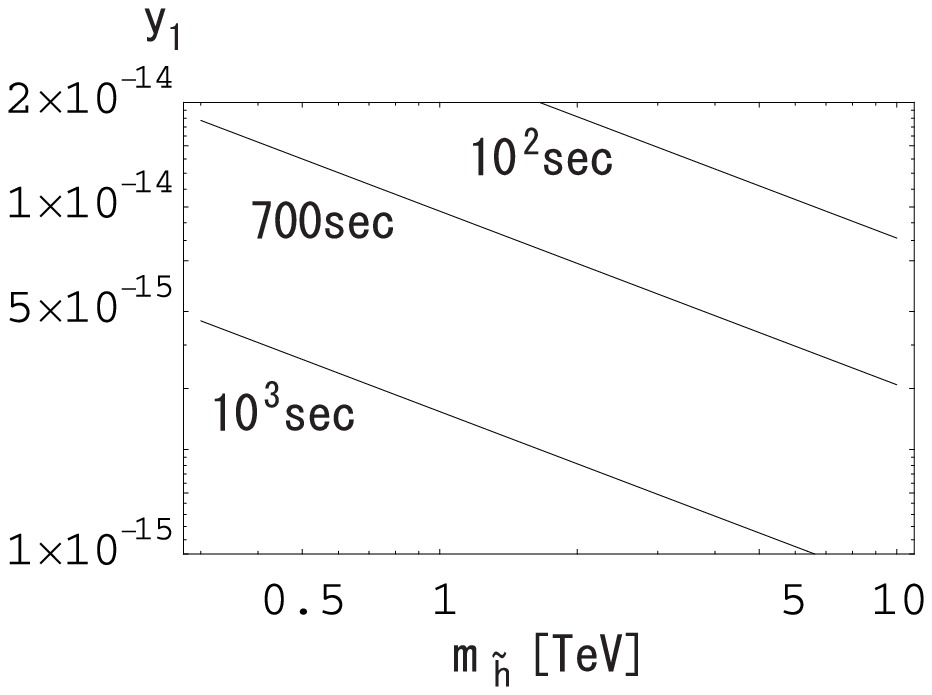}
\caption{The left figure shows the dependence of $1/\Gamma_{\tilde{\nu}_L\rightarrow\phi\phi}$ on the left-handed
 sneutrino mass and $\lambda_1$, where $m_D=10^{-3}$ eV is taken. The region lower than the bold horizontal line,
 which corresponds $\lambda_1\simeq4\times10^{-9}$, is allowed. The right figure shows 
$1/\Gamma_{\tilde{h}\rightarrow\tilde{\nu}_L\psi}\simeq1/\Gamma_{\tilde{h}\rightarrow\nu_L\phi}\simeq y_1^2m_{\tilde{h}}/32\pi$.}
\label{fig:1}
\end{figure}

We consider two body decay processes of the sneutrino and the higgsino in this section. A peculiar decay process 
of the sneutrino in the MaVaNs scenario is $\tilde{\nu}_L\rightarrow\phi\phi$, whose decay rate is given by 
 \begin{eqnarray}
  \Gamma_{\tilde{\nu}_L\rightarrow\phi\phi}\simeq\frac{\lambda_1^2m_D^2}{16\pi m_{\tilde{\nu}_L}}.
 \end{eqnarray}
When we 
take $m_{\tilde{\nu}_L}=1$TeV, $\lambda_1=10^{-9}$ and $m_D=10^{-3}$eV, we have 
$1/\Gamma_{\tilde{\nu}_L\rightarrow\phi\phi}\simeq3\times10^{22}$sec. A larger left-handed sneutrino mass or 
smaller $\lambda_1$ and $m_D$ suppress this process. The dependence of 
$1/\Gamma_{\tilde{\nu}_L\rightarrow\phi\phi}$ on the left-handed sneutrino mass and $\lambda_1$ is shown in 
Fig.\ref{fig:1}.

The higgsino decay is another peculiar one in this scenario. There are two important processes, 
$\tilde{h}\rightarrow\nu_L\phi$ and $\tilde{\nu}_L\psi$. The decay rates of these processes are given by
 \begin{eqnarray}
  \Gamma_{\tilde{h}\rightarrow\nu_L\phi}\simeq\frac{y_1^2m_{\tilde{h}}}{32\pi},\hspace{5mm} 
  \Gamma_{\tilde{h}\rightarrow\tilde{\nu}_L\psi}
  \simeq\frac{y_1^2(m_{\tilde{h}}^2-m_{\tilde{\nu}_L}^2)^2}{32\pi m_{\tilde{h}}^3}.
 \end{eqnarray} 
When we take $y_1=10^{-14}$ and $m_{\tilde{h}}=1$TeV, $1/\Gamma_{\tilde{h}\rightarrow\nu_L\phi}\simeq 700$ sec. 
If we assume that the sneutrino mass is less than the higgsino mass but the same size, 
$1/\Gamma_{\tilde{h}\rightarrow\tilde{\nu}_L\psi}$ is the same order as 
$1/\Gamma_{\tilde{h}\rightarrow\nu_L\phi}$. Larger $m_{\tilde{h}}$ or $y_1$ enhances these processes as shown in 
Fig.\ref{fig:1}. The above processes as we discussed may affect on many LSP cold dark matter models. Such effects
 of the Mass Varying Neutrinos scenario will be discussed elsewhere.

\section{Summary}
We presented a supersymmetric Mass Varying Neutrinos model and detailed analyses of the vacuum structure of the 
scalar potential taking account of the finite imaginary part of the acceleron. The potential has four local 
minima and the dark energy is identified with the false vacuum energy. A metastable vacuum are realized near the 
origin in field space, $\phi_1\simeq0$ and $\phi_2=0$, due to the supersymmetry breaking effect. This metastable 
vacuum is enough stable in the age of the universe. In the model, the observed dark energy scale of order 
$(10^{-3}\mbox{eV})^4$ is understood by gravitationally suppressed supersymmetry breaking scale, 
$F(\mbox{TeV})^2/M_{\mbox{{\scriptsize Pl}}}$.

Following these considerations, the case with non-vanishing vacuum expectation values of sneutrinos have been 
discussed. We have found that the VEVs of sneutrinos are strictly constrained such as 
$\langle\tilde{\nu}_{L,R}\rangle<10^{-39}$ eV in order that terms with respect to sneutrinos in the acceleron 
potential do not spoil the model. This means that the vanishing or tiny VEVs of sneutrinos are favored in the 
proposed supersymmetric MaVaNs model. Finally, two body decay processes $\tilde{\nu}_L\rightarrow\phi\phi$, 
$\tilde{h}\rightarrow\nu_L\phi$ and $\tilde{h}\rightarrow\tilde{\nu}_L\psi$ are discussed. 
The obtained decay rates are $1/\Gamma_{\tilde{\nu}_L\rightarrow\phi\phi}\simeq3\times10^{22}$ sec. and 
$1/\Gamma_{\tilde{h}\rightarrow\nu_L\phi}\simeq1/\Gamma_{\tilde{h}\rightarrow\tilde{\nu}_L\psi}\simeq 700$ sec.. 
 Effects on many LSP cold dark matter model from these processes is expected to provide 
interesting physics.

\section*{Acknowledgments}
We would like to thank Takehiko Asaka and Koichi Hamaguchi for helpful discussions. The work of R.T. has been 
supported by the Japan Society of Promotion of Science. M.T. has been also supported in part by scientific grants
 from the Ministry of Education, Science, Sports, and Culture of Japan (No. 17540243 and 19034002).

\end{document}